\documentclass[fleqn,10pt]{wlscirep}
\usepackage[utf8]{inputenc}
\usepackage[T1]{fontenc}
\usepackage{bm}
\usepackage{tcolorbox}
\usepackage{multicol}
\newenvironment{Figure}
  {\par\medskip\noindent\minipage{\linewidth}}
  {\endminipage\par\medskip}

\title{Simulating Topological Order on Quantum Processors}

\author[1,2,*]{Adam Gammon-Smith}
\author[3,4]{Michael Knap}
\author[3,4]{Frank Pollmann}

\affil[1]{School of Physics and Astronomy, University of Nottingham, Nottingham, NG7 2RD, UK}
\affil[2]{Centre for the Mathematics and Theoretical Physics of Quantum Non-Equilibrium Systems, University of Nottingham, Nottingham, NG7 2RD, UK}
\affil[3]{Technical University of Munich, TUM School of Natural Sciences, Physics Department, Lichtenbergstr. 4,85748 Garching, Germany}
\affil[4]{Munich Center for Quantum Science and Technology (MCQST), Schellingstr. 4, 80799 M{\"u}nchen, Germany}

\affil[*]{e-mail: adam.gammon-smith@nottingham.ac.uk}

\begin{abstract}
It is an ongoing quest to realize topologically ordered quantum states on different platforms including condensed matter systems, quantum simulators and digital quantum processors. Unlike conventional states characterized by their local order, these exotic states are characterized by their non-local entanglement. The consequences of topological order can be as profound as they are surprising, ranging from the emergence of fractionalized anyonic excitations to potentially providing a scalable platform for quantum error correction. This deep connection to quantum computing naturally motivates the realization and study of topologically ordered quantum states on quantum processors. However, due to the non-local nature of these states, their study presents a challenge for near-term quantum devices. This Perspective aims to review the recent progress towards the experimental realization of topologically ordered quantum states, their potential applications, and promising directions of future research. 
\end{abstract}
\begin{document}

\flushbottom
\maketitle

\thispagestyle{empty}

\noindent \textbf{Key points:}

\noindent \textbf{Website summary:}

\section*{Main}
Quantum many-body systems can host exotic phases of matter that go beyond the paradigm of conventional symmetry-breaking. Instead, they are characterized by patterns of long-range entanglement and non-local order~\cite{Wen2004,wen2017zoo}. These topological phases exhibit remarkable properties, including ground state degeneracy that depends on system topology, robust edge modes protected by global symmetries, long-range quantum entanglement, intrinsic resilience to local noise, and the emergence of anyonic excitations—emergent particles that defy conventional bosonic or fermionic statistics~\cite{Kitaev2003, Nayak2008}. As a result, they have become a central topic of interest in condensed matter physics and quantum information science. Box~1 provides an overview of different classes of topological matter.

Despite their theoretical appeal and practical potential—especially for fault-tolerant computation\cite{Kitaev1997}—realizing and characterizing topologically ordered states remains an outstanding challenge. Their non-local nature makes them elusive to standard experimental probes, and requires high level of control over quantum systems. Key obstacles include decoherence, engineering complex entanglement structures, and devising scalable state preparation and verification protocols. Nevertheless, important progress has been made in identifying and probing topological phases. A landmark example is the observation of the fractional quantum Hall effect (FQHE), which provided the first experimental realization of a topologically ordered phase with anyonic quasiparticles~\cite{tsui1982, laughlin1983}. More recently, symmetry-protected topological (SPT) phases have been realized in cold atom systems~\cite{Endres2011, Hilker2017,deLeseleuc2019,Sompet2022, su2025}, and quantum spin liquid behavior has been investigated in frustrated magnetic materials~\cite{Balents2010, Savary2016, Knolle19}. While many results are still not definitive, these efforts have significantly advanced our understanding of non-trivial quantum phases.

A complementary and increasingly powerful approach involves using digital quantum processors to simulate and explore topologically ordered systems. These platforms allow for the controlled preparation, manipulation, and measurement of entangled quantum states, enabling direct access to properties that are difficult to probe in conventional condensed matter experiments. Gate-based quantum processors are especially promising for emulating the intricate entanglement patterns that underpin topological order. However, a key challenge remains: How to efficiently harness current Noisy Intermediate-Scale Quantum (NISQ) hardware to study these complex quantum phases?

In this Perspective, we review recent progress toward the realization of topological quantum states using programmable quantum processors. We focus on both symmetry-protected topological (SPT) phases~\cite{Gu2009,Pollmann2010,Chen2011} and intrinsically topologically ordered (TO) phases~\cite{Wen1990}; see Box~1 for definitions and distinctions. We examine state preparation techniques, tools for detecting non-local order parameters, and methods to image emergent anyonic excitations. We highlight key experimental milestones across diverse quantum platforms, including superconducting qubits, trapped ions, Rydberg atom arrays, and others. Finally, we outline promising future directions, such as enhancing preparation fidelity, integrating error-correcting protocols, and developing diagnostics for topological order in noisy settings. A deeper understanding of these exotic phases not only furthers fundamental physics but also brings us closer to realizing robust quantum technologies rooted in the principles of topology.

\begin{figure}[!t]
\begin{tcolorbox}[colback=blue!5, colframe=white, arc=0mm, size=small]
\vspace{7pt}
\section*{Box 1}
\vspace{-7pt}
\noindent\rule{\textwidth}{0.5pt}

\vspace*{+10pt}
{\LARGE Topological Phases of Matter}\\
\noindent\rule{\textwidth}{0.25pt}
\begin{multicols}{2}

The conventional Landau paradigm classifies phases of matter through symmetry breaking and associated local order parameters. However, this framework fails to capture a broad class of phases known as topological phases, which are not characterized by local order but instead by non-local entanglement and global topological features (see Fig.~\ref{fig: phase diagram}).\cite{wen2017zoo}
Examples of topological phases include the Haldane phase in spin chains~\cite{Haldane1983}, the fractional quantum Hall effect (FQHE)~\cite{tsui1982,laughlin1983}, and the toric code model~\cite{Kitaev2003}. These phases can be broadly categorized into two types: phases with short-range entangled \emph{Symmetry-Protected Topological} (SPT) order, and with long-range entangled \emph{intrinsic topological} order.

\begin{Figure}
\centering
\includegraphics[width=0.8\columnwidth]{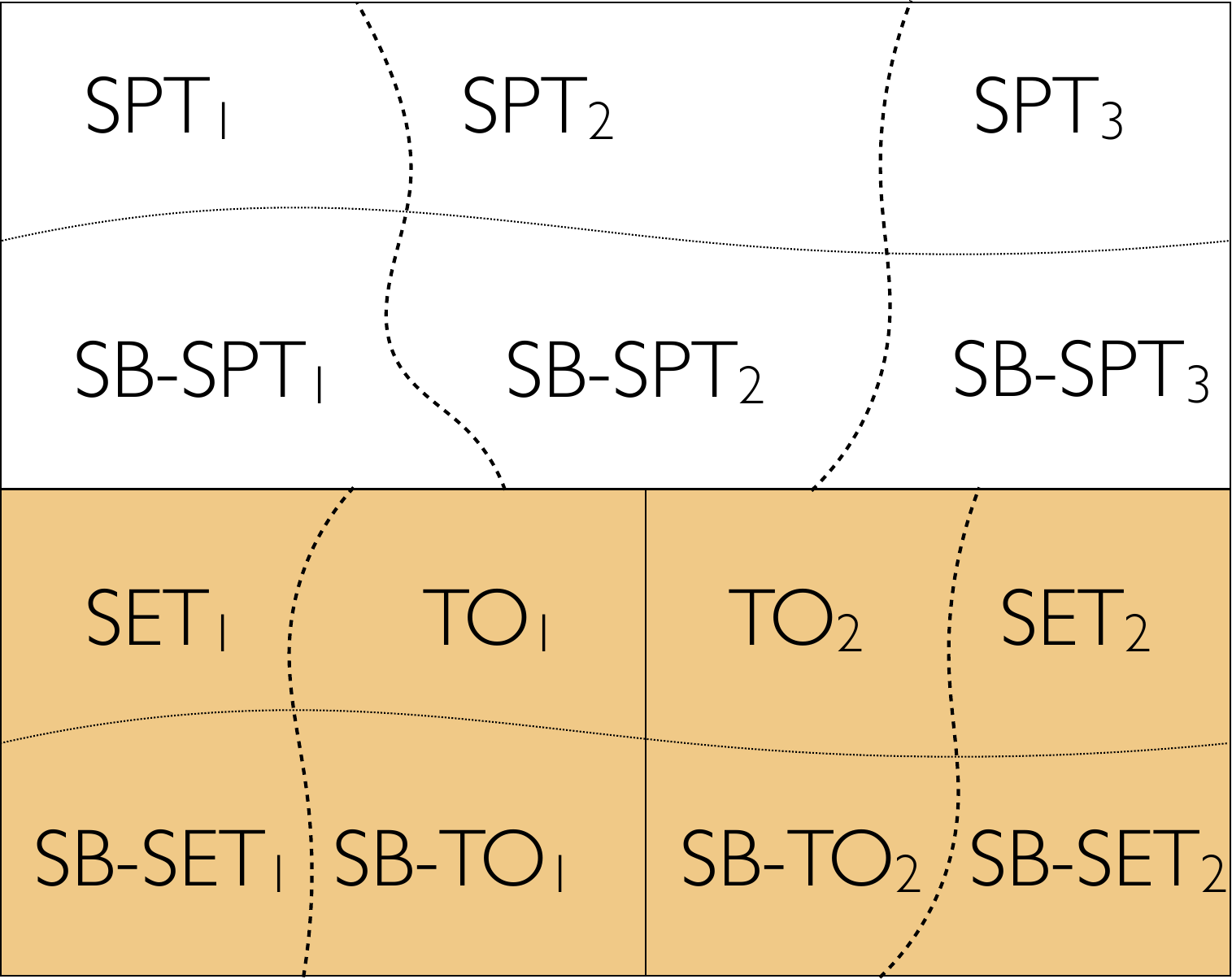}\\
\caption{\footnotesize Quantum many-body phase diagrams split into the trivial phase (white) and different Topologically Ordered (TO) phases (orange). When symmetries are enforced, the phase diagram is additionally split into different Symmetry Protected Topological (SPT) and Symmetry Enriched Topological  (SET) phases (separated by dashed lines) and Symmetry Broken (SB) phases (separated by dotted lines).}
\label{fig: phase diagram}
\end{Figure}
\hphantom{x}\\[-4pt]
{\bf Symmetry Protected Topological (SPT) Order}

SPT phases are short-range entangled and trivial in the absence of symmetry, but they exhibit nontrivial behavior when specific symmetries are enforced.\cite{Gu2009,Pollmann2010,Chen2011} 
A canonical example in one dimension is the celebrated Haldane phase of spin-1 chains, which exemplifies a bosonic SPT phase protected by spin rotations, time-reversal, or inversion symmetry.

\hphantom{x}\\[-4pt]
\emph{Edge or surface modes}---One dimensional SPT phases, like the Haldane phase\cite{Haldane1983}, are characterized by degenerate or gapless edge modes that cannot be removed without breaking the symmetry or closing the bulk gap and by degeneracies in the entanglement spectrum\cite{Pollmann2010}.

\hphantom{x}\\[-4pt]
\emph{Non-local order parameters}---SPT order cannot be distinguished by any local order parameter. Instead they are characterized by non-local order parameters\cite{denNijs1989,Pollmann2012} such as a string order parameter in the case of the Haldane phase.

\hphantom{x}\\[-4pt]
SPT phases do not exhibit fractionalized bulk excitations or ground state degeneracy on closed manifolds. Their classification relies on mathematical tools such as group cohomology, cobordism theory, and generalized cohomology.

\hphantom{x}\\[-4pt]
{\bf Intrinsic Topological Order}

Intrinsic topological order refers to phases that are robust to any local perturbations and do not depend on symmetries for their stability. The FQHE is the prototypical example of such a phase~\cite{tsui1982,laughlin1983}, and its theoretical understanding has been furthered by topological quantum field theories and category theory.

\hphantom{x}\\[-8pt]
\emph{Anyons}---One of the characteristic features of topological order is the presence of emergent excitations called anyons~\cite{Nayak2008}. 
These fractionalized excitations can be treated like particles with properties distinct from bosons and fermions. 
Anyons can be distinguished by their statistics, including exchange statistics differing from $\pm 1$. Certain phases of matter can host non-abelian anyons, whose statistics has a non-commuting matrix structure~\cite{Nayak2008}. 
This property forms the basis for topological quantum computation~\cite{Kitaev1997}.

\hphantom{x}\\[-8pt]
\emph{Long-range entanglement and grounds state degeneracy}---Topologically ordered systems cannot be distinguished by any local order parameter. Instead they are characterised by long-range entanglement, which is a system size independent negative contribution to the entanglement entropy,  referred to as topological entanglement entropy~\cite{Levin2006,Kitaev2006b}. This long-range entanglement also results in a ground state degeneracy that depends on the boundary conditions.

\hphantom{x}\\[-8pt]
\emph{Symmetry-enriched topological} (SET) order refers to topologically ordered phases where the presence of global symmetries further distinguishes and enriches the structure of the phase.

\end{multicols}
\vspace{4pt}
\label{fig:box}
\end{tcolorbox}
\end{figure}

\begin{figure}[!t]
\centering
\includegraphics[width=.95\linewidth]{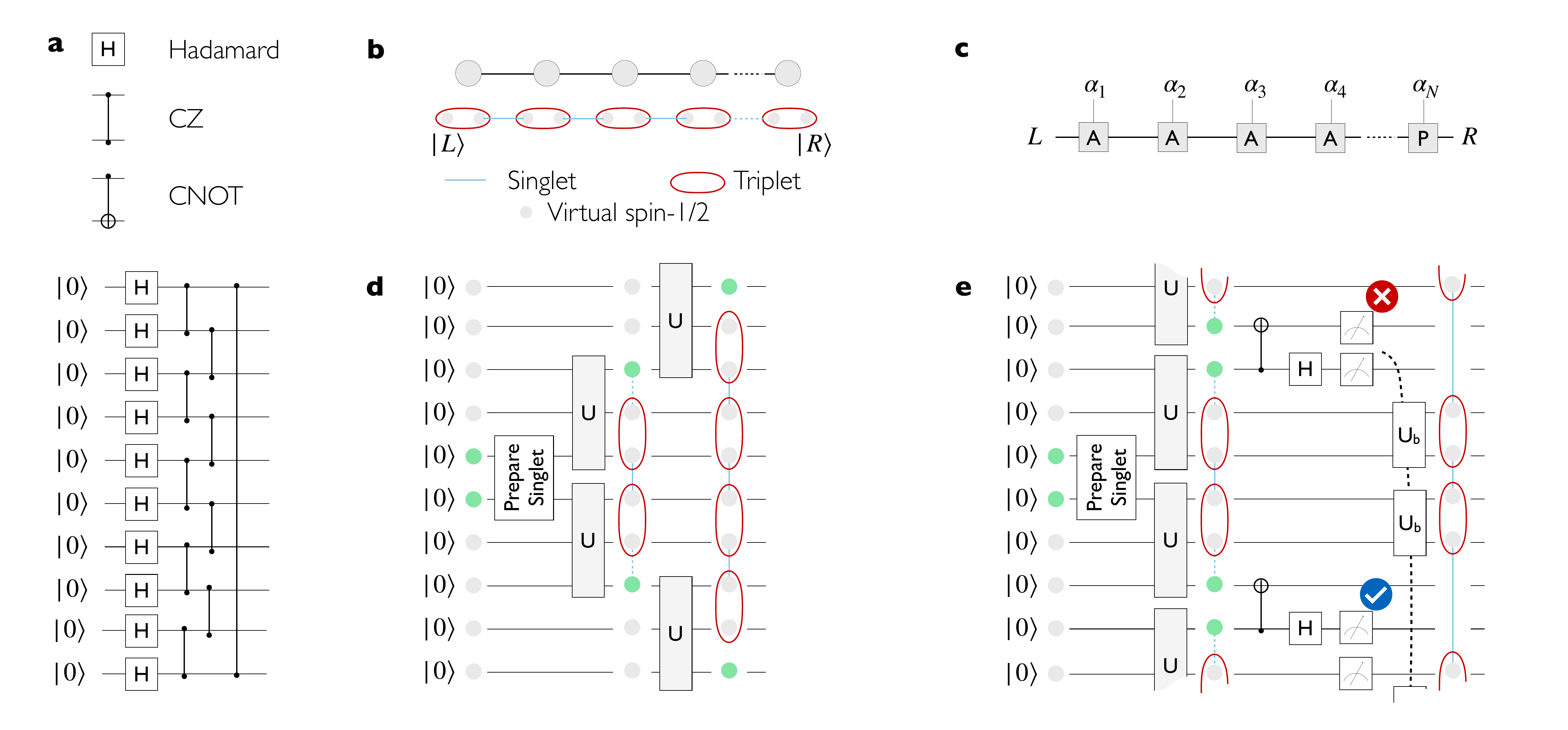}
\caption{Preparation of SPT states on quantum computers: (a) Unitary circuit for preparing the cluster state \cite{choo2018measurement}.
(b) Illustration of the spin-1 AKLT state using two projected spin-1/2 particles.
(c) Diagram of the MPS representation of the AKLT state, with boundary conditions denoted by $L$ and $R$, respectively \cite{Smith2023}.
(d) Sequential unitary preparation of the AKLT state using a linear-depth circuit \cite{Smith2023}.
(e) Circuit diagram for measurement-assisted preparation \cite{Smith2023}.}
\label{fig: SPT preparation}
\end{figure}

\section*{Symmetry Protected Topological (SPT) States}

SPT phases are quantum phases of matter that arise only when certain symmetries are preserved. Unlike conventional phases distinguished by spontaneous symmetry breaking, SPT phases remain invariant under the protecting symmetry and cannot be smoothly connected to trivially disordered phases without breaking this symmetry, see Box~1.

\subsection*{State preparation techniques for SPT phases}

To illustrate the key concepts of SPT phases and the different approaches used for the state preparation, we begin by introducing two prominent models that exemplify SPT order: the cluster state and the Affleck-Kennedy-Lieb-Tasaki (AKLT)\cite{Affleck1987} model.   
\paragraph{Cluster model:}We begin with the one-dimensional (1D) cluster model, whose ground state exhibits SPT order, yet remains straightforward to implement on gate-based quantum computers. 
The Hamiltonian is given by

\begin{equation}
H_{\text{cluster}} = -\sum_i Z_{i-1} X_i Z_{i+1} \label{eq:cluster_hamiltonian},
\end{equation}
where $\{X, Y, Z\}$ represent the Pauli matrices and the individual terms $Z_{i-1} X_i Z_{i+1}$ are known as stabilizers, as they are mutually commuting. 
The cluster state and also its higher-dimensional versions have been considered prominently in the context of measurement-based quantum computation \cite{Raussendorf2001,Jiang2025}.
The 1D cluster state has zero correlation length and exhibits SPT order \cite{Pollmann2010, Chen2011,Son2011}. 
The SPT phase is protected by several symmetries, including the $\mathbb{Z}_2 \times \mathbb{Z}_2$ symmetry, which enforces the conservation of the following parity operators $P_{\text{odd}} = \prod_i X_{2i+1}$ and $P_{\text{even}} = \prod_i X_{2i}$.
Another protecting symmetry is the combination of time-reversal symmetry (i.e., complex conjugation) together with a global $\mathbb{Z}_2$ parity symmetry. 
By using unitary gates that break the protecting symmetries, the cluster state can be prepared on a quantum computer using a very simple constant depth circuit involving two steps starting from $\otimes_i |0\rangle$ as illustrated in Fig.~\ref{fig: SPT preparation}a:  (i) Apply Hadamard gates to all qubits. (ii) Apply two consecutive layers of controlled-$Z$ gates.
For the case of open boundary conditions,  the first and last qubits are not connected, leading to a symmetry-protected fourfold ground state degeneracy.
This simple protocol makes the cluster state an ideal test case for the preparation and characterization of SPT phases on quantum computers. 
Note that the exact representation of the state by a finite-depth circuit is possible due to its zero correlation length. More general SPT states do not possess this property, which motivates the following example.

\paragraph{AKLT Model:} The AKLT model describes a one-dimensional (1D) spin-1 chain. Unlike the cluster model, its ground state exhibits a finite correlation length. Moreover, it is adiabatically connected to the ground state of the spin-1 Heisenberg model, which is relevant to a variety of 1D quantum magnetic systems\cite{Buyers1986}.
The Hamiltonian includes both bilinear and biquadratic nearest-neighbor interactions:

\begin{equation}
H = \sum_{i} \vec{S}_i \cdot \vec{S}_{i+1} + \frac{1}{3} (\vec{S}_i \cdot \vec{S}_{i+1})^2.
\end{equation}

Originally introduced in the context of Haldane's conjecture of the Heisenberg model exhibiting a finite energy gap for integer spin \cite{Haldane1983}, the AKLT model remains a subject of significant interest.
In particular, it serves as a paradigmatic example for a frustration-free Hamiltonian with an exact matrix product state (MPS) ground state, referred to as the (spin-1) AKLT state.
The SPT order of the AKLT state~\cite{Pollmann2010, Chen2011} has been demonstrated to be protected by any of the following symmetries: time-reversal symmetry, spatial reflection symmetry, or dihedral $\mathbb{Z}_2 \times \mathbb{Z}_2$ symmetry (which corresponds to local $\pi$ rotations about two orthogonal axes at each site). 
As a consequence, the AKLT state exhibits key characteristics of an SPT phase, such as fractionalized symmetries at the edges, and a hidden string order \cite{denNijs1989, Pollmann2012}. 
Moreover, the AKLT state has been shown to be a resource state for measurement based quantum computing \cite{Else2012}. 

The spin-1 AKLT state can be understood in terms of virtual spin-1/2 pairs as shown in Fig.~\ref{fig: SPT preparation}b, alternating between singlet and triplet configurations, the latter corresponding to physical spin-1 sites. 
The state can be expressed as an MPS of the form
\begin{equation}
|\Psi\rangle = \sum_{\vec{m}} \langle L| A^{m_1} A^{m_2} \dots A^{m_{N-1}} A^{m_N} |R\rangle |\vec{m}\rangle,
\end{equation}
as illustrated in Fig.~\ref{fig: SPT preparation}c.
The physical indices $m_i \in \{+,0,-\}$ enumerate the spin-1 triplet states and the virtual degrees of freedom reflect the entanglement of the singlet states. The tensors are given by  
$ A^- = -\sqrt{2/3} \, \sigma^{-} $, $ A^0 = -\sqrt{1/3} \, \sigma^z$, and  $A^+ = \sqrt{2/3} \, \sigma^{+}$.
Thus, the AKLT state is an MPS with bond dimension $\chi=2$ and physical dimension $d=3$.
The boundary choices $\langle L|$ and $|R \rangle$ project onto $S=1/2$ states, reflecting the fractionalized edge states $\{\uparrow,\downarrow\}$, characteristic of SPT phases.

There are several routes to preparing symmetry protected AKLT state on a quantum computer.
Note that in contrast to the cluster state, the finite correlation length of the AKLT state prevents an exact representation in terms of a (simple) finite depth unitary circuit.
The conceptually simplest preparation of the AKLT state is by directly converting the MPS into a sequential circuit\cite{Schoen2007,Smith2023} which will generically require a unitary circuit with a depth linear in system size, as illustrated in Fig.~\ref{fig: SPT preparation}d.
At each time step, the site preparation unitary $U$ acts on two qubits forming a spin-1 site (gray circles) and a memory qubit (green circles), which propagates correlations between spin-1 sites. 
The unitaries $U$ can be directly derived from the tensors $A^{m_i}$.
To minimize circuit depth, the approach exploits spatial inversion symmetry, growing the AKLT state simultaneously from both ends with two parallel memory qubits.
Note that by using connections to MPS of infinite systems, it is also possible to measure observable in a formally infinite state on a system using a finite number of qubits \cite{Smith2022}. 
In a similar spirit, a connection between MPS and quantum channels allows a ``holographic'' simulation of the 1D state using a number of qubits that increases logarithmically with the virtual dimension.\cite{Foss-Feig2021}

An alternative approach to preparing the AKLT state utilises mid-circuit measurements and classical feedback in conjunction with finite-depth unitary circuits~\cite{Piroli2021,Smith2023}.
This type of operation has been demonstrated on a number of different experimental platforms. 
The idea is to create local parts of the system, and then to couple them through projective measurements. The measurement outcome will be probabilistic, with one of the outcomes being desired, and the others indicating some kind of error. For certain classes of states, the outcomes of these measurements can then be fed back to determine a finite depth circuit to correct these errors. This approach has also been applied to the AKLT state, \cite{Smith2023} as shown in Fig.~\ref{fig: SPT preparation}e. This measurement and feedback approach is more generally motivated by a connection between MPS and quantum circuits. If the MPS tensors have a particular type of symmetry, then the errors can be corrected locally and passed down the chain using a finite depth circuit, allowing for a finite depth preparation of the state. It is understood that this method realises a strict sub-class of MPS~\cite{}. To access more general states, there are adiabatic procedures~\cite{Wei2023}, as well as approximate schemes for translating MPS to quantum circuits~\cite{Malz2024}.

\subsection*{Detection and characterization of SPT phases}

\begin{figure}[t]
\centering
\includegraphics[width=1\linewidth]{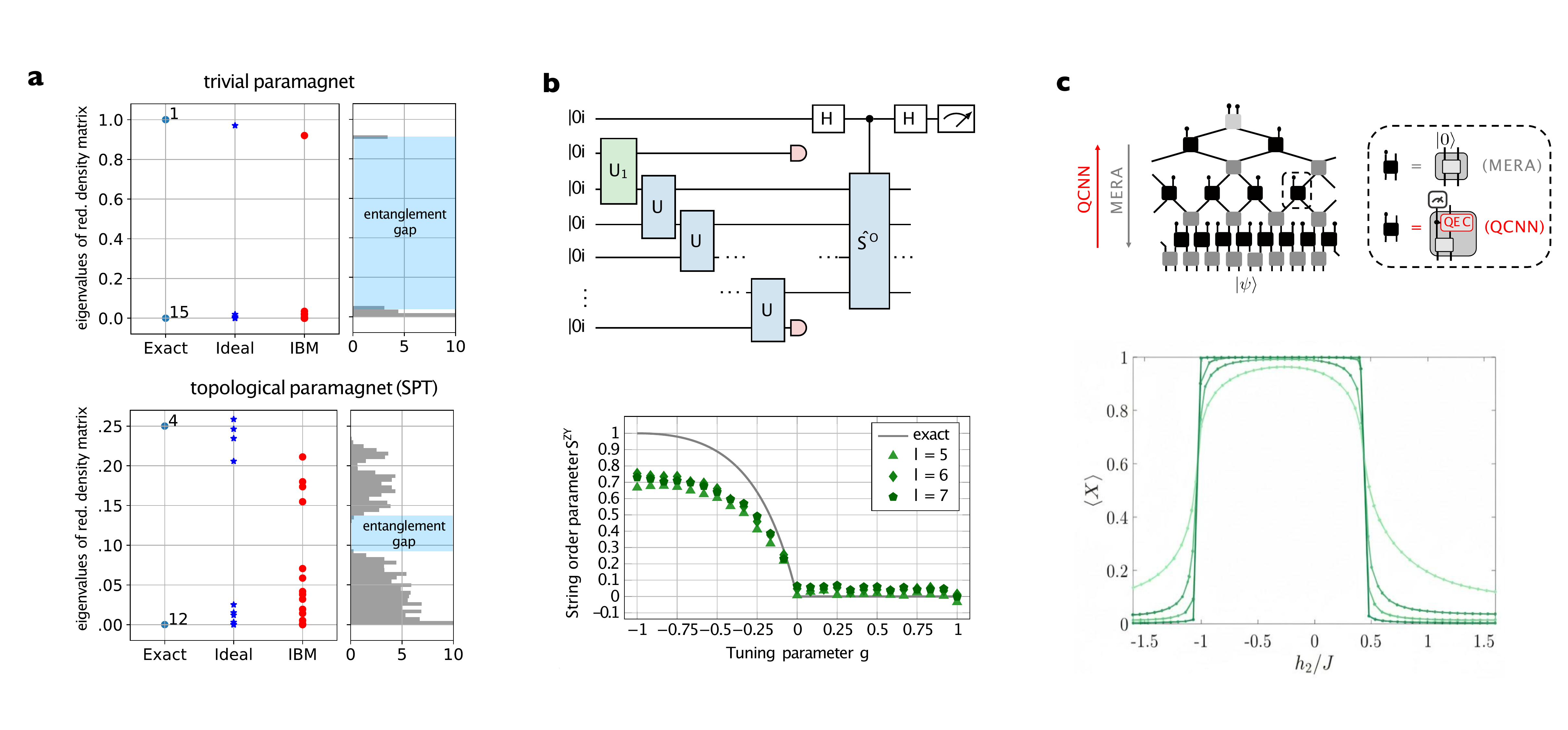}
\caption{Characterizing SPT order on quantum computers: 
(a) Entanglement spectrum of the cluster state measured on an IBM quantum computer (red) compared to exact values (light blue). The entanglement gap and near-degenerate low-lying levels signal topological order, though statistical noise (blue) and experimental imperfections lift the expected degeneracy.\cite{choo2018measurement}
(b) Detection of the SPT-to-trivial phase transition via a string order parameter measured using interferometry on an IBM quantum computer.\cite{Smith2022}
(c) Exact QCNN output along a path crossing an SPT phase obtained using a QCNN.\cite{Cong2019}
}
\label{fig: spt measurement}
\end{figure}

Given an SPT ground state that has been prepared on a quantum computer, we require access to non-local properties of the wave function to correctly characterize and verify the states. 
One of the first approaches to detecting SPT phases on quantum computers involved measuring the entanglement spectrum \cite{choo2018measurement}, shown in Fig.~\ref{fig: spt measurement}a, where characteristic degeneracies serve as indicators of SPT order \cite{Pollmann2010}. 
However, this approach relies on access to the reduced density matrix, which requires full-state tomography---an experimentally demanding process that scales exponentially with the size of the subsystem.
Another approach is to evaluate string-order parameters, whose exact form depend on the specific representation of the protecting symmetry \cite{denNijs1989,Pollmann2012}.
For spin chains, these take the form of Pauli strings with non-local support. 
Strictly speaking, the order parameter is the infinite length limit of these non-local operators, although a long finite length string can provide an accuracy exponential in the length of the string. 
To measure these operators directly requires measuring a large number of qubits, corresponding to the non-trivial support of the string. 
This leads to exponential compounding of measurement errors, which may be one of the dominant sources of error in NISQ devices. 
Alternatively, it is possible to use interferometric measurement to extract the value from ancilla qubits. This allows one to trade measurement error for coherent gate errors and decoherence, see Fig.~\ref{fig: spt measurement}b. 

Inspired by advances in machine learning, quantum convolutional neural networks (QCNNs) as shown in Fig.~\ref{fig: spt measurement}c have been explored as powerful tools for identifying and measuring non-local string order parameters in SPT phases~\cite{Cong2019}.
For the classification an analytic solution was found for a particular solvable phase diagram, and the optimisation of unitary gates was demonstrated more generally.\cite{Cong2019} Furthermore, classically trained QCNNs and support vector machines (SVMs) have been implemented on quantum processors.~\cite{Herrmann2022,Sadoune2024}
More recently, through the use of known fixed point states (such as the cluster state), it was possible to map out phase diagrams by learning on a random set of states generated from these fixed points by applying random symmetry preserving circuits.\cite{Liu2023}
These circuits ensure that the state remain in the same phase as the fixed point, but allowed the model to efficiently sample the space in a model agnostic manner. 
%
In this way this approach is generalized beyond specific models, since only the form of the symmetry was required.

\section*{Topologically Ordered (TO) States}

In contrast to SPT phases, topologically ordered (TO) phases exhibit a more intrinsic form of topological order, robust against any local perturbation, independent of symmetry~\cite{Simon2023}. This remarkable robustness arises from their defining characteristics: a ground state degeneracy that depends on the topology of the system, long-range entanglement, and the emergence of exotic anyonic excitations (see Box~1 for a detailed comparison).  These remarkable properties make TO phases a promising platform for fault-tolerant quantum computation, in which quantum information is inherently protected from local noise.

\begin{figure}[t]
\centering

\includegraphics[width=\linewidth]{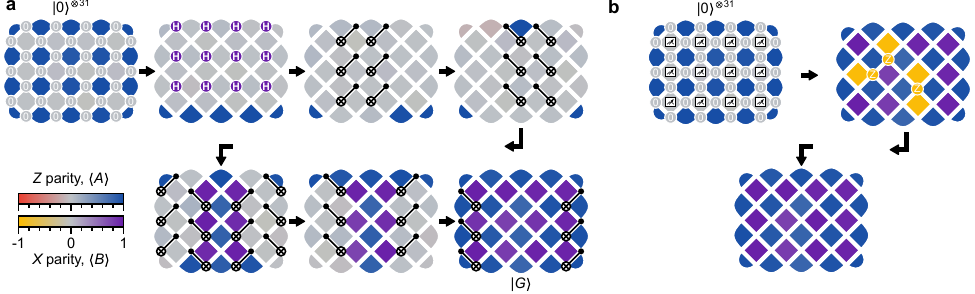}
\caption{Preparing topological order on quantum processors. (a) Unitary quantum circuit to prepare the toric code ground state, along with measurements of the star and plaquette stabilisers at each step.~\cite{Satzinger2021} The state can be realised using a single layer of Hadamard gates and a linear sequence of nearest neighbour CNOT gates. (b) Finite depth preparation of the toric code using adaptive quantum circuits with measurement and feedback.\cite{Kitaev2003} Starting with a product state that satisfies the stars, measuring the plaquette operators randomly projects onto the $\pm 1$ eigenstates. Any remaining plaquette defects can be removed by a single layer of single qubit gates, which are determined by the measurement outcomes.}
\label{fig: TC Google}
\end{figure}

\subsection*{Realizing Topologically Ordered States on Quantum Hardware}

Realizing topologically ordered states on quantum hardware requires precise control over quantum systems. We review the fundamental concepts and techniques employed to engineer these complex states, highlighting the challenges and recent progress in this area. 

Quantum many-body phases can be characterized either by their Hamiltonians or by their quantum states. Two gapped Hamiltonians belong to the same phase if they can be continuously connected by a path of gapped Hamiltonians. Similarly, two states are in the same phase if a finite-depth quantum circuit can transform one into the other~\cite{Chen2010c}. A trivial phase and \emph{topologically ordered} phase, or distinct topologically ordered phases, therefore cannot be interconverted by such transformations. There have also been recent efforts to generalise this circuit definition of topological ordered pure states to density matrices~\cite{Coser2019,Shengqi2024,Sohal_2025,Wang_2025,Ellison_2025}, inspiring constructions of topological order that survive at finite temperature~\cite{Zhou2025}. Symmetry constraints on the Hamiltonians or circuits lead to additional symmetry-protected topological and symmetry-broken phases, as illustrated in Fig.~\ref{fig: phase diagram}. It is important to note that topological order does not exist in one dimension. As a concrete example, we introduce the toric code, which provides a clear illustration of topologically ordered states and their properties.

\paragraph{Toric code:} The toric code, introduced by Alexei Kitaev~\cite{Kitaev1997,Kitaev2003}, is a paradigmatic model in the study of topological order due to its simplicity and ability to capture many key features of topologically ordered systems. It is a lattice model of spin-1/2 degrees of freedom arranged on a square lattice (and others), as shown in Fig.~\ref{fig: TC Google}. The toric code Hamiltonian is the sum of commuting stabilizers
\begin{equation}\label{eq: toric code}
H = -J_A \sum_{+} A_+ - J_B \sum_{\square}B_\square \qquad\text{with}\qquad  A_+ =  \prod_{j\in+} Z_j, \qquad  B_\square=\prod_{j \in \square} X_j .
\end{equation}
Here the $A_+$ and $B_\square$ terms represent ``star'' and ``plaquette'' operators, respectively, which enforce local constraints on the spins. These stabilisers commute with each other, and a ground state $|\psi\rangle$ is left unchanged by them, i.e., $A_+ |\psi\rangle = |\psi\rangle$ and $B_\square|\psi\rangle = |\psi\rangle$ for all stars and plaquettes. The fact that these stabilizer terms commute with each other allows for an exact solution and makes the topological properties of the model more accessible. Despite its simplicity, the toric code elegantly captures many general features of topologically ordered systems. The exact solvability of the model allows us to simply describe the ground state(s). They can be viewed as equal superpositions of closed loops of flipped spins in the Z-basis. The closed loops satisfy the star terms, while the plaquette operators connect different loop configurations, leading to a superposition of configurations in the ground states. However, if we put the system on a torus there are closed loops around the two periodic directions, which cannot be generated by local terms such as those in the Hamiltonian, leading to the fourfold degeneracy of the ground state associated with the winding parity in those two directions. 

Preparing the toric code ground state presents a challenge: it cannot be achieved from a simple product state using a unitary quantum circuit with only a constant number of layers. Instead, creating this state requires circuits whose complexity scales linearly with the system size~\cite{Satzinger2021}, although logarithmic scaling is possible with parallel implementation of long-range gates~\cite{Koenig2009}. Recently, researchers have successfully demonstrated preparation of the toric code ground state using unitary quantum circuits on a superconducting quantum processor~\cite{Satzinger2021} (see Fig.~\ref{fig: TC Google}a). 

The toric code represents  a specific type of topological order. Other examples include non-abelian topological order. String-net models~\cite{Levin2005} offer microscopic spin Hamiltonians capable of realizing diverse examples of this more complex form of topological order.  Furthermore, recent theoretical proposals for preparing string-net ground states with linear-depth quantum circuits have been experimentally validated for the double Fibonacci model~\cite{Liu2022, xu2024nonabelian,Minev2025}.

Similarly to the progress in SPT state preparation, recent work has demonstrated that many topologically ordered states can be efficiently created using finite-depth unitary circuits augmented with measurements and classical feedback~\cite{Piroli2021,Tantivasadakarn2023a,Tantivasadakarn2023b}. This approach leverages a process known as ``gauging,'' where an initial state is prepared using a relatively shallow circuit. Subsequent measurements in a specific basis reveal defects, which are then corrected locally via feedback operations. For example, the toric code can be initialized in a simple polarized product state satisfying star operators, followed by measurement of plaquette operators and single-qubit corrections for any resulting defects (see Fig.~\ref{fig: TC Google}b). A step toward an experimental measurement-based realization of the toric code has been achieved in Ref.~\cite{Bluvstein2022}, in which measured defects have been corrected classically. 
While this approach has the ability to prepare a wide variety of topological orders---including non-Abelian phases and even potentially universal ones~\cite{Lo2025}---it remains an open question whether all topologically ordered states, particularly those hosting Fibonacci anyons, can be realized using finite-depth adaptive circuits. Notably, it has been shown that a large class of non-Abelian topological orders based on both solvable and non-solvable groups can be prepared using adaptive circuits with depth scaling logarithmically with system size.~\cite{Lu2022}

Crucially, verifying the successful creation of these topologically ordered states requires characterizing their fundamental properties, one of which is the degeneracy of their ground state. The number of degenerate ground states in topologically ordered systems is intimately linked to the system’s boundary conditions. Open boundaries typically lead to a unique ground state, but introducing ‘rough’ boundaries can induce degeneracy---a property exploited in quantum error correction schemes~\cite{Bravyi1998,Dennis2002,Satzinger2021,Google2025}. Periodic boundary conditions, such as those imposed on a torus, result in a degeneracy equal to the number of emergent anyon types. In the toric code, this manifests as a four-fold ground state degeneracy.   Recent experiments have successfully prepared and verified all 22 degenerate ground states in the non-abelian $D_4$ topological order.~\cite{Iqbal2024}

\begin{figure}[t]
\centering
\includegraphics[width=\linewidth]{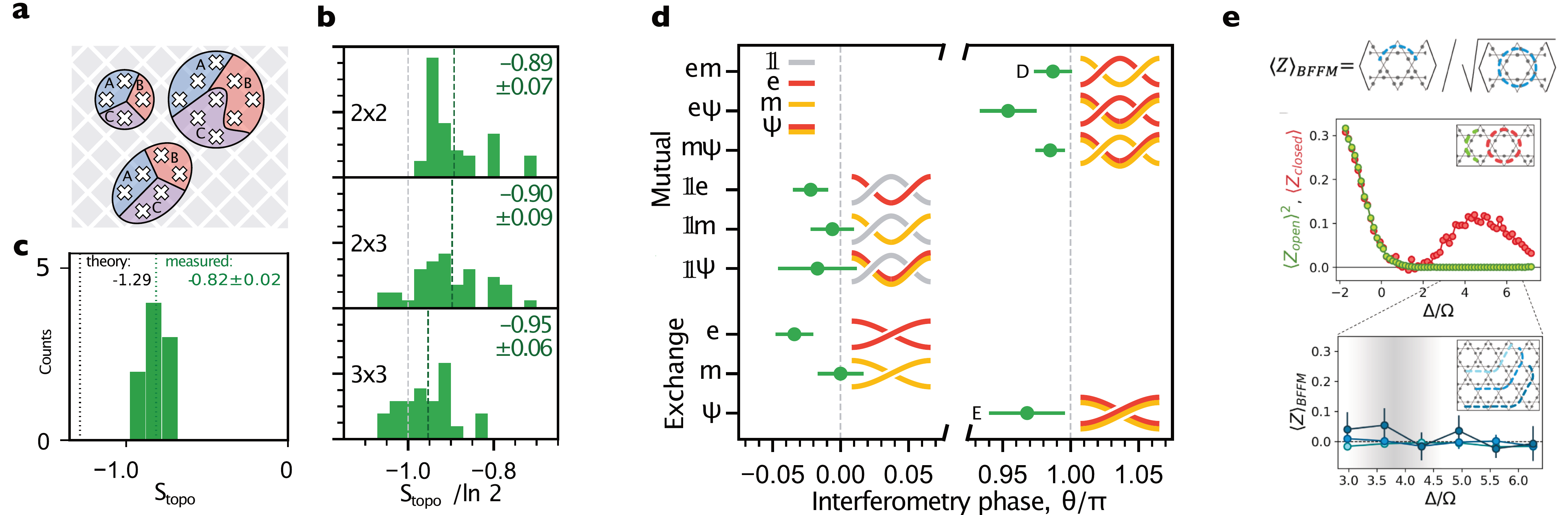}
\caption{Characterising topological order. (a) Different size partitions for computing the topological entropy on a quantum processor~\cite{Satzinger2021}. (b) Experimental data corresponding to these partitions for the toric code, computed using full state tomography and randomized measurement methods.~\cite{Satzinger2021} (c) Topological entanglement entropy measurement for the double Fibonacci model on a superconducting quantum processor.~\cite{xu2024nonabelian} (d) Measured anyon statistics in the toric code.~\cite{Satzinger2021} (e) Measurement of the Fredenhagen-Marcu order parameter for a system of Rydberg atoms indicating $\mathbb{Z}_2$ topologically order.~\cite{Semeghini2021}
}
\label{fig: TO experiment}
\end{figure}

\subsection*{Topological entanglement entropy}

A defining characteristic of topologically ordered (TO) phases is their long-range entanglement. This entanglement manifests as a constant contribution to the entanglement entropy when measuring a subsystem. Specifically, consider a disk-shaped region in a TO state with boundary of length $l$. The entanglement entropy $S$ for this region takes the form
\begin{equation}
    S = \alpha l - \gamma + ... \qquad  \text{where} \;\; \gamma = \log \mathcal{D}
\end{equation}
where $\alpha$ is a non-universal coefficient that depends on the microscopic details along the boundary, and all remaining terms vanish in the limit $l \rightarrow \infty$. 
The topological entanglement entropy (TEE), given by $\gamma=\log\mathcal{D}$ where $\mathcal{D}=\sqrt{\sum_i d_i}$, represents a non-negative contribution that reflects the long-range entanglement present in topologically ordered states. This quantity is independent of microscopic details and provides a signature of topological order.
It represents the total quantum dimension associated with the underlying topological order, reflecting a deep connection between ground state properties and emergent anyonic excitations.\cite{SHI2020168164}

Isolating the topological entanglement entropy $\gamma$, involves strategic partitioning schemes designed to cancel non-universal contributions. One approach involves dividing into four disk-like regions $A,B,C$ and $D$ arranged such that $A$, $B$, and $C$ share a triple intersection and $D$ is the complement (as shown in Fig.~\ref{fig: TO experiment}a).\footnote{Refs.~\cite{Kitaev2006b,Levin2006} also include an alternative arrangement with disconnected regions.} The topological entropy~\cite{Kitaev2006b,Levin2006} can than be extracted via
\begin{equation}
    S_\text{topo} = S_{ABC} - S_{AB} - S_{AC} - S_{BC} + S_{A} + S_{B} + S_{C} = -\gamma,
\end{equation}
where $S_{AB}$ is the entanglement entropy of the combined region $A \cup B$, and similarly for the other terms. The idea is that non-universal contributions along the boundaries cancel out, leaving only the topological contribution, $S_\text{topo} = -\gamma$, which provides a clear signature of topological order.

Measuring entanglement entropy experimentally is a formidable challenge as the required number of measurements scales exponentially with the size of the subsystem. Despite these obstacles, progress has been made in measuring TEE using superconducting quantum devices. For the abelian $\mathbb{Z}_2$ topological order in the toric code~\cite{Satzinger2021}, a series of experiments to determine the TEE has been successfully implemented. The experiment involved preparing the ground state of the toric code on a superconducting quantum processor and then measuring entanglement entropy using both full state tomography and techniques based on randomized measurements across different partitions of the system. By carefully analysing these measurements, good agreement has been found with the expected value $S_\text{topo} = -\ln 2$ (see Fig.~\ref{fig: TO experiment}b). The TEE of the non-abelian double Fibonacci model has been measured as well~\cite{xu2024nonabelian}. The expected value for this model is $S_\text{topo} = -\ln(1+\varphi^2)$, where $\varphi = (1+\sqrt{5})/2$ is the golden ratio. While achieving precise measurements in this case is challenging due to the increases circuit depth required to prepare these states, the experiment produced results that were consistent with the theoretical prediction. Specifically, the bound $-\ln(1+\varphi^2) < S_\text{topo}^\text{measured} < -\ln 2$ has been found (see Fig.~\ref{fig: TO experiment}c). These experimental findings demonstrate the ability to measure non-local topological properties of states on quantum processors.

\subsection*{Anyons in topologically ordered systems}
 
Anyons are emergent point-like excitations, which behave neither like bosons nor fermions. They are characterized by their \emph{statistics}, which can be revealed through experiments where anyons are moved around each other, forming different braids in spacetime. The exchange statistics of anyons is the phase acquired when two indistinguishable anyons are exchanged. For abelian anyons, this phase is a simple (complex) number (e.g., $\pm 1$), while for non-abelian anyons, the phase depends on the fusion channel and can be represented by matrices. In two dimensions, moving one anyon around another in a loop also acquires a non-trivial phase---known as mutual statistics.

For topologically ordered systems there exist Wilson string operators that move the excitations around the system. At the fixed points such as the toric code, these Wilson strings take a particularly simple form of Pauli strings of either $Z$ or $X$ operators. Applying Wilson strings that exchange particles or drag one around the other induces the corresponding phases ($\pm 1$). To extract this phase in Ref.~\cite{Satzinger2021} a Hadamard test was used, where the Wilson string was implemented as a controlled operation with an ancilla, and the phase was extracted by performing tomography only on the ancilla qubit as shown in Fig.~\ref{fig: TO experiment}d. For the toric code, this reveals an emergent fermion, along with $e$ and $m$ anyons that have $-1$ mutual statistics. The emergence of fermionic excitations in the toric code has also lead to proposals for local mappings from fermions to qubit in two dimensions.~\cite{Bravyi2002, Verstraete_2005, Derby2021,Chen2023} Recent experiments on trapped ion quantum computers have demonstrated the practical benefit of these local mappings over more commonly used non-local mappings such as Jordan-Wigner transformations.~\cite{Nigmatullin2024}

Away from fixed-point states such as the toric code, Wilson string operators will not have such simple forms and instead will be unknown dressed operators with exponentially decaying support. These ideal Wilson operators can nonetheless be used to identify topologically ordered states through the Fredenhagen-Marcu (FM) order parameter.~\cite{Fredenhagen1983} The FM order parameter involves taking the ratio of expectation values of open and closed string operators (see Fig.~\ref{fig: TO experiment}e). In topologically ordered phases, this ratio should be exactly zero because open string operators create excitations at their ends, while dividing by the square root of the expectation value of the closed operator normalizes for the decay due to unknown string operators. Ref.\cite{Semeghini2021} measured the FM order parameter in trapped Rydberg atoms, indicative of topologically ordered states in the 
$\mathbb{Z}_2$ phase (see Fig.\ref{fig: TO experiment}e). Recently, there has also been increased theoretical interest in using the FM order parameter to study topological phase transitions.~\cite{Xu2024-key} Determining the FM order parameter in the vicinity of the topological phase transition, requires resolving an exponentially small denominator and numerator, rendering it hard to determine experimentally. An alternative strategy based on quantum error correction has been proposed to overcome this limitation in large portions of the phase diagram.~\cite{Liu2025}

While the toric code hosts abelian anyons, topologically ordered systems can also support non-abelian anyonic excitations. These are characterized by multiple fusion channels. For example, Fibonacci anyons exhibit two possible outcomes when fused: either they combine into a trivial topological charge (1) or another Fibonacci anyon ($\tau$). This is analogous to the spin singlet and triplet states for pairs of spin-1/2 particles. Additionally, the result of braiding non-abelian anyons depends on the fusion channel, meaning that their braiding statistics are described by matrices rather than phases. Ref.~\cite{xu2024nonabelian} used a variation of the unitary String-Net construction from Ref.~\cite{Liu2022} to realize the doubled Fibonacci model on a superconducting quantum computer and measure the effects of non-abelian braiding. When certain braids were performed, the configuration of anyons remained invariant, but they measured a superposition of fusion channels as expected for these anyons. A similar process was applied in a variant of the toric code with lattice defects. These defects can host excitations with non-abelian statistics (specifically Ising anyons), which were recently measured~\cite{andersen2023nonabelian}. In order to demonstrate the unique nature of non-abelian anyons, Ref.~\cite{Iqbal2024} considered a more complex procedure involving three anyons, which could be viewed as a braid in space-time corresponding to Borromean rings. These rings have the property that any pair of rings is unlinked, and it is only when all three are involved that they are linked. For abelian models, this braid can always be trivially unlinked. However, an experiment on a trapped ion quantum computer measured a phase of $-1$, indicating the non-abelian nature of the realised states.~\cite{Iqbal2024}

\section*{Outlook}%
With quantum processors allowing controlled preparation and manipulation of exotic topological states, a new experimental frontier has opened up at the intersection of condensed matter, quantum information, and computer science.
While early demonstrations focused on simple and mostly fined tuned model systems, we are now witnessing a shift towards exploring the dynamics and stability of these phases. 
Scaling simulations and at the same time preserving fidelity remains a key challenge, requiring not only improved error mitigation but also circuit constructions that preserve topological invariants under noise. 

The relevance of topological order in the context of fault tolerant quantum processing is particularly compelling. 
While topological order forms the bedrock of many promising quantum error correction schemes, such as the toric code, recent demonstrations of active error correction with increasing code distance~\cite{Google2025,Bluvstein2024} highlight both progress and remaining hurdles. 
Scaling to larger number of qubits needed for even modest fault-tolerant universal computation 
will require not only hardware improvements but also innovations in protected gate implementations. 
Beyond the toric code, theoretical research is expanding the landscape of potential topological codes – including colour codes~\cite{Bombin2006}, other abelian and non-abelian models~\cite{Koenig2010,Wooton2014,Dauphinais2019,Zhu2020,Schotte2022}, higher-dimensional constructions~\cite{Dennis2002,Bravyi2013}, and quantum low-density parity check (qLDPC) codes~\cite{Gottesman2014,Breuckmann2021} (see the ``Error Correction Zoo'' \cite{ErrorCorrectionZoo}). 
Recent experiments have started to explore  braiding and fusion properties of non-Abelian anyons \cite{xu2024nonabelian,andersen2023nonabelian,Iqbal2024}, but developing complete topological gate sets and resilient encoding schemes are critical next steps. The development of completely new types of quantum error correction schemes can be envisioned as well. Further progress could turn these exotic theoretical concepts into practical tools for fault-tolerant quantum information processing.~\cite{Pachos2012,Nayak2008}

Recent quantum simulations largely focussed on fixed-point wavefunctions such as the toric code or doubled Fibonacci model. 
A crucial next step is extending these capabilities to explore more general, perturbed phases of topological order, including those lacking exact analytical solutions. 
This requires developing methods capable of characterizing topological properties in non-fixed point states, encompassing both variational approaches such as tensor network inspired quantum algorithms, as well as techniques for directly probing emergent behaviour in these more complex systems. 
The development of robust order parameters that can be efficiently measured on noisy intermediate-scale quantum (NISQ) devices is particularly important. 
Recently, quantum circuits for realizing  fracton topological order have been constructed and transition between distinct symmetry enriched topological phases have been analysed.\cite{Boesl2025}
An emerging frontier with respect to the stability of topological orders lies in understanding topological order at finite temperatures~\cite{Hastings2011,Castelnovo2008,Yoshida2011} and exploring phases that offer passive error protection.~\cite{Knill2000,Brown2016,Placke2024,Li2024}  
Many conventional topological phases are destroyed by thermal fluctuations, necessitating active stabilization. However, recent theoretical work suggests the possibility of intrinsically stable, finite-temperature topological phases with reduced reliance on continuous measurement, potentially simplifying the requirements for fault tolerance. These finite temperature topologically ordered phases are also of fundamental interest in condensed matter physics.

An important conceptual question in quantum many-body physics is how to classify states that can be transformed into one another via low-depth quantum circuits supplemented by local operations and classical communication. \cite{Piroli2021} In one-dimensional systems with Abelian symmetries, it has been shown that all matrix product states (MPS) become equivalent under shallow symmetric circuits combined with symmetric measurements and feed forward, effectively collapsing the distinctions between symmetry-protected topological (SPT) phases.\cite{gunn2025phases} In contrast, for non-Abelian symmetries, some SPT phases remain distinct, illustrating that symmetry and measurement fundamentally reshape the landscape of phase classification.

Real-time dynamics offer another exciting avenue, where quantum computers have the potential to excel over classical systems by managing substantial entanglement growth. In that vain, dynamics of anyons and strings has been investigated.~\cite{Cochran2024} Early efforts have further shown that quantum platforms can effectively investigate the dynamics of fractionalized excitations and their behaviour under periodic driving. These driven quantum systems can exhibit novel topological properties that are uniquely non-equilibrium, such as the transmutation of anyons.\cite{Po2017,Will2025} The landscape of non-equilibrium phases with topological order is largely unexplored thus far and so is how to probe and utilize this type of topological order. Analysing these aspects are exciting future directions.

Quantum computers are not only becoming practical platforms for encoding and manipulating topological quantum matter—they are emerging as laboratories for discoveries. 
As experiments push beyond fixed-point wavefunctions into dynamical, noisy, and even finite-temperature regimes, the theoretical boundaries of topological order are being redefined. 
This evolving synergy between hardware, theory, and computation promises not just advances in quantum technology, but a deeper understanding of  entangled quantum phases.

\bibliography{references}

\section*{Acknowledgements}

A.G-S.~was supported by the UK Research and Innovation (UKRI) under the UK government’s Horizon Europe funding guarantee [grant number EP/Y036069/1]. M.K. and F.P. acknowledge support from the Deutsche Forschungsgemeinschaft (DFG, German Research Foundation) under Germany’s Excellence Strategy–EXC–2111–390814868, TRR 360 – 492547816, FOR 5522 (project-id 499180199),  and DFG grants No. KN1254/1-2, KN1254/2-1, the European Research Council (ERC) under the European Union’s Horizon
2020 research and innovation programme (grant agreement No 851161), the European Union (grant agreement No 101169765), as well as the Munich Quantum Valley, which is supported by the Bavarian state government with funds from the Hightech Agenda Bayern Plus.




\end{document}